\documentclass[prd%
               ,floatfix%
               ,preprintnumbers%
               ,superscriptaddress%
               ,amsmath%
               ,amssymb%
           ,nofootinbib
               ]{revtex4}
\usepackage[dvips]{color,graphicx}% Include figure files
\usepackage{dcolumn}% Align table columns on decimal point
\usepackage{bm}% bold math
\definecolor{ref}{rgb}{1.0,0.0,0.0}
  \makeatletter
  \def\mathcomposite{%
     \@ifstar
        {\def\@mathcomposite@option{%
            \baselineskip\z@skip\lineskiplimit-\maxdimen}%
         \@mathcomposite}%
        {\let\@mathcomposite@option\offinterlineskip
         \@mathcomposite}}
  \def\@mathcomposite{%
     \@ifnextchar[\@@mathcomposite{\@@mathcomposite[0]}}
  \def\@@mathcomposite[#1]#2#3#4{%
     #2{\mathchoice
        {\@mathcomposite@{#1}{#3}{#4}\displaystyle{1}}%
        {\@mathcomposite@{#1}{#3}{#4}\textstyle{1}}%
        {\@mathcomposite@{#1}{#3}{#4}%
         \scriptstyle\defaultscriptratio}%
        {\@mathcomposite@{#1}{#3}{#4}%
         \scriptscriptstyle\defaultscriptscriptratio}}}
  \def\@mathcomposite@#1#2#3#4#5{%
     \vcenter{\m@th\@mathcomposite@option
        \dimen@\f@size\p@\dimen@#1\dimen@\dimen@#5\dimen@
        \divide\dimen@ 18
        \edef\@mathcomposite@skipamount{\the\dimen@}%
        \ialign{\hfil$#4##$\hfil\cr
           #2\crcr
           \noalign{\vskip\@mathcomposite@skipamount}%
           #3\crcr}}}
  \makeatother

\def\bfm#1{\mbox{\boldmath $#1$}}
\def\bfsm#1{\mathstrut\mbox{\scriptsize{\boldmath $#1$}}\mathstrut}

\newcommand{\ep}{\varepsilon}

\newcommand{\beq}{\begin{equation}}
\newcommand{\eeq}{\end{equation}}
\newcommand{\ba}{\begin{array}}
\newcommand{\ea}{\end{array}}
\newcommand{\bea}{\begin{eqnarray}}
\newcommand{\eea}{\end{eqnarray}}
\newcommand{\bal}{\begin{align}}  % align in amsmath is better than eqnarray
\newcommand{\eal}{\end{align}}
\newcommand{\bi}{\begin{itemize}}  %\setlength{\itemsep}{0\parsep}}
\newcommand{\ei}{\end{itemize}}
\newcommand{\ben}{\begin{enumerate}}  %\setlength{\itemsep}{0\parsep}}
\newcommand{\een}{\end{enumerate}}

\newcommand\hide[1]{}

\newcommand{\tr}{\mbox{tr}}

\newcommand{\Det}{\mbox{Det}}

\renewcommand{\Re}{{\rm Re}\,}
\renewcommand{\Im}{{\rm Im}\,}
\newcommand{\ie}{{i.e.}}

\newcommand{\ds}[1]{
  \setbox0=\hbox{\ensuremath{#1}}
  \hbox to\wd0{\hbox to0pt{\hbox to\wd0{\hss/\hss}\hss}\box0}}

\begin{document}
\title{On the neutrality issue in the Polyakov-loop NJL model}
\author{H. Abuki}\email{abuki@th.physik.uni-frankfurt.de}
\affiliation{Institut f\"ur Theoretische Physik, J.W. Goethe
Universit\"at, D-60438 Frankfurt am Main, Germany}
\author{M. Ciminale}\email{marco.ciminale@ba.infn.it}
\affiliation{I.N.F.N., Sezione di Bari, I-70126 Bari, Italy}
\affiliation{Universit\`a di Bari, I-70126 Bari, Italy}
\author{R. Gatto}\email{raoul.gatto@physics.unige.ch}
\affiliation{D\'epartement de Physique Th\'eorique, Universit\'e
de Gen\`eve, CH-1211 Gen\`eve 4, Switzerland}
\author{M. Ruggieri}\email{marco.ruggieri@ba.infn.it}
\affiliation{I.N.F.N., Sezione di Bari, I-70126 Bari, Italy}
\affiliation{Universit\`a di Bari, I-70126 Bari, Italy}
\date{\today}

\begin{abstract}
We elucidate how the color neutrality is harmed in the Polyakov
 Nambu-Jona Lasinio (PNJL) model at finite density within the adopted mean field approximation. Also we point out
how usual assumption about the diagonal form of the Wilson loop may fail in
 the presence of the diquark condensate on several grounds.
\end{abstract}
\maketitle
\section{Introduction}
In recent years several aspects of the Quantum Chromodynamics (QCD) phase diagram in the $\mu-T$ plane have been
investigated. Here $\mu$ denotes the quark chemical potential and $T$ corresponds to the temperature. The studies are
mainly related to the breaking and/or restoring of the global symmetries of the QCD lagrangian in the ground state. The
high temperature-small chemical potential region of the phase diagram can be accessed by lattice calculations, see for
example\cite{Aoki:2006br,Schmidt:2006us,Philipsen:2005mj,Heller:2006ub} and reference therein, as well as effective
models, among others chiral perturbation theory (ChPT) \cite{Gasser:1983yg,Gasser:1984gg}, quark meson coupling (QMC)
\cite{Guichon:1987jp,Saito:1994ki} and the Nambu-Jona Lasinio model(NJL) \cite{Nambu:1961tp,Nambu:1961fr} . On the
other hand the dual region of low temperature-large chemical potential is not accessible nowadays by lattice
calculations because of the well known sign problem of 3-color QCD (for reviews
see\cite{Ejiri:2004yw,Splittorff:2006vj,Splittorff:2006fu}). In this case our knowledge of the QCD phase diagram is
mainly based on phenomenological models at intermediate density and perturbative QCD at very high density.

One of the most interesting features of the QCD phase diagram at
high temperature and low density is the chiral crossover. The QCD
chiral symmetry is explicitly broken by the bare quark masses and
spontaneously broken by chiral condensates\cite{Nachtmann:1978zh}.
In the vacuum, i.e. at $\mu=T=0$, the chiral condensate
$\langle\bar\psi \psi\rangle$ is expected to be of the order of
$-(250~\text{MeV})^3$. Lattice calculations show that there exists a
temperature of the order of $200$ MeV at which the chiral condensate
smoothly drops down to a lower value, and at the same time chiral
susceptibilities develop a peak, signaling the existence of a smooth
crossover. The Goldstone bosons related to the chiral symmetry
breaking are the pions, which are massive below the phase transition
temperature because of the bare quark masses. Their mass increases
above the critical temperature so that they decouple from the low
energy spectrum, and this can be interpreted as a further signal of
the restoration of the chiral symmetry. The smooth crossover becomes
a true phase transition when one artificially sets the bare quark
masses to zero, so that the chiral symmetry becomes an exact
symmetry of the QCD lagrangian, and above the critical temperature
the chiral condensate is zero.

Lattice calculations show that in the neighborhood of the chiral crossover a Polyakov loop crossovers
occurs\cite{Banks:1983me,Karsch:1994hm,Aoki:1998wg}. The Polyakov loop $\Phi$ is an order parameter for deconfinement
in the pure gauge theory\cite{Polyakov:1978vu,Susskind:1979up,Svetitsky:1982gs}. In more detail, the expectation value
of the Polyakov loop $\langle\Phi\rangle$ is related to the free energy $F_q$ of a heavy (thus non dynamical) quark by
the relation $\langle\Phi\rangle\propto\exp(-\beta F_q)$. The calculations of $\langle\Phi\rangle$ as a function of the
temperature in lattice-QCD (LQCD) give the result that $\langle\Phi\rangle = 0$ for $T< T_0$ and $\langle\Phi\rangle
\neq 0$ at $T > T_0$. Here $T_0$ denotes the so called deconfinement temperature of the pure glue theory, $T_0 \approx
270$ MeV. The LQCD results imply that for $T<T_0$ the free energy of a static quark is infinite and therefore it can
not be produced. On the other hand for $T>T_0$ the expectation value of the Polyakov loop is no longer zero, thus the
free energy of a static quark is finite and it can be produced. Beside this the correlation of a Polyakov loop and a
Polyakov antiloop at finite distance, namely $\langle\bar\Phi(0)\bar\Phi(x)\rangle$, is related to the interaction
energy of a heavy quark and a heavy antiquark. At $T<T_0$ the correlation grows linearly with the distance $x$, thus
enforcing the confining feature of the low temperature region, while at $T>T_0$ it becomes a screened Coulomb
potential. The Polyakov loop is a gauge singlet but it is not invariant under an element of the $Z_3$ group. Therefore
in the deconfined phase the $Z_3$ symmetry is spontaneously broken while in the confined phase it is unbroken. For this
reason the Polyakov loop is often referred to as the order parameter of the confinement-deconfinement transition.

In presence of dynamical quarks one can no longer regard $\Phi$ as an order parameter. The very reason for this is that
dynamical quarks break explicitly the $Z_3$ symmetry and thus one expects $\langle\Phi\rangle \neq 0$ even in the
previously called confined phase. Nevertheless the Polyakov loop is supposed to serve as an indicator of the phase
transition, since LQCD results show that a Polyakov loop crossover occurs almost in coincidence with the chiral
crossover at high temperature. The coincidence of the two crossovers suggests that chiral symmetry restoration and
deconfinement are intimately related.

These results are confirmed by effective model calculations. Among
the numerous effective models  the most celebrated one is certainly
the NJL model (for reviews see
\cite{Klevansky:1992qe,Hatsuda:1994pi,Buballa:2003qv}). In the NJL
model one replaces the gluon-mediated quark interaction by a
4-fermion interaction, and the chiral condensate is computed by
solving a non linear gap equation mimicking the BCS effective model
of ordinary superconductors\cite{Bardeen:1957mv}. Once the
parameters of the model are adjusted in order to reproduce some
phenomenological vacuum properties of QCD (meson masses, chiral
condensate and pion decay constant) the model can be used to predict
the existence of a chiral crossover at a temperature of the order of
$200$ MeV, in agreement with LQCD. Unfortunately the original
version of the NJL model does not incorporate gluons as dynamical
degrees of freedom. As a consequence the interplay between the
chiral and the Polyakov loop crossover can not be discussed within
the model. This obstacle has been recently surmounted by the NJL
model with the Polyakov loop (PNJL model in the
following)\cite{Fukushima:2003fw,Meisinger:1995ih}.

In the PNJL model one introduces an effective potential $\cal U$ for the Polyakov loop and its conjugate. Its
parameters are tuned to reproduce the LQCD results with no dynamical quarks, namely the deconfinement temperature $T_0$
and the thermodynamics of the pure glue liquid for temperatures close to $T_0$. Moreover assuming the existence of a
background temporal gluon field, the coupling of the Polyakov loop to the dynamical quarks is achieved via the QCD
coupling of the background field and the quarks themselves. For more formal details see next Section. The PNJL model
successfully reproduces several LQCD results at zero and small quark chemical potential, as well as results with
imaginary chemical potential.

We have demonstrated in \cite{Abuki:2008ht}, that the Polyakov Nambu-Jona Lasinio (PNJL) model~\cite{Fukushima:2003fw}
{\em only} at nonzero baryon chemical potential may have some intrinsic problem of color neutrality. As stressed there,
this issue does not exist at zero density, in no contradiction with the zero density results in \cite{Hell:2008cc}.
However, once we introduce the baryon chemical potential, any PNJL-type model is no longer free from the color
neutrality problem. This may be viewed as an artifact of the PNJL model, or possibly of the mean-field approximation
scheme usually taken in the literature. In this article we wish to clarify how this problem can emerge at finite
density, irrespective to the existence of diquark fields.

\section{Formal settings}
First we illustrate the problem at finite density but without the
diquark condensate. To be confident, we here start with the NJL model
for two-flavor quarks, combined with a matrix model in order to derive
the PNJL type model~\cite{Ghosh:2007wy}. Apart from the gluon part, the
partition function of the PNJL model is given as
\beq
\ba{rcl}
  Z[\mu,T]&=&\int%
  \Pi_{\bfsm{x}}dL(\bfm{x})\,%
  \int dq d\bar{q}%
  e^{-\int_0^{1/T} d\tau\int dx \mathcal{L}_E(q,\bar{q};A_4)},
\ea
\eeq
with the Euclidean lagrangian density defined as
\beq
 \mathcal{L}_E(q,\bar{q},A_4)=q^+(\gamma_0(\partial_\tau+i A_4(x))%
  +i\bfm{\gamma}%
  \cdot{\nabla}+m-\ds{\mu})q-G\left[%
  (\bar{q}q)^2+(\bar{q}i\gamma_5\bfm{\tau}q)^2\right].
\label{lag} \eeq $L$ is the Wilson-loop matrix, $L\in\mathrm{SU}(3)$. In the hypothesis that the system is
well-described by the mean-field approximation at large distance \cite{Dumitru:2005ng}, we replace the above group
integral over the all space with a single integral over $L$, \ie, we take $L(\bfm{x})$ to be a constant over the whole
space, which is an implicit assumption also in the PNJL model \cite{Fukushima:2003fw}. \beq
 L(\bfm{x})=P_{\tau}e^{i\int_0^{1/T} d\tau
 A_4(x)}\stackrel{\mathrm{PNJL}}{\to}L=e^{iA_4/T},
\eeq where $A_4=\sum_{\alpha=1}^{8}\phi_\alpha\lambda_\alpha$ with $\{\phi_\alpha\}$ the real parameters. By
introducing the auxiliary field $\sigma\sim 2G\bar{q}{q}$, and then integrating out the quark fields we obtain \beq
 Z[\mu,T]%
  =\int %
  \!dL\!\int_{\mathrm{P.B.}}{\mathcal D}\sigma(x)%
  e^{-S_{\mathrm{eff}}[A_4;\sigma(x)\mu,T]},
\eeq
where the integral over $L$ is replaced by the global one. The effective
action in the presence of constant $A_4$ background is thus
\beq
  S_{\mathrm{eff}}[A_4,\sigma(x);\mu,T]=\int_0^{1/T}%
  d\tau\int\frac{\sigma(x)^2}{4G}-\log\Det_{x,y}%
  \left[(-i\ds{D}_x[\mu,A_4]%
  +m-\sigma(x))\delta_{\mathrm{P}}(x-y)\right],%
\eeq
where the covariant derivative is defined by
$i\ds{D}[\mu,A_4]\equiv\gamma_0(\partial_\tau+iA_4-\mu)%
-i\bfm{\gamma}\cdot\nabla_x$, and $\delta_{\mathrm{P}}(x-y)$
is the delta function periodic in imaginary time $\tau$.
In the above equation we have assumed that in the ground state pion
condensation does not take place, resulting in the vanishing of the
pseudoscalar condensate $\langle\bar{q} \gamma_5 \bm{\tau} q\rangle$.
This simplification is justified by our previous results that
show that in the physical limit pion condensation is forbidden in the
ground state of the model~\cite{Abuki:2008tx,Abuki:2008nm,Abuki:2008wm}.

We look for the homogeneous ground state via a saddle-point
approximation. In this case, the approximated action is space-time
independent and equal to $\beta V\Omega[A_4,\sigma;\mu,T]$, with
thermodynamic potential $\Omega$ given by
\beq
\ba{rcl}
 \Omega(L,\sigma;\mu,T)=\frac{\sigma^2}{4G}%
 -2N_{\mathrm{f}}T\int\frac{d\bfsm{p}}{(2\pi)^3}%
 \tr_c\ln\left(1+L^+e^{-(E-\mu)/T}%
 \right)%
  -2N_{\mathrm{f}}T\int\frac{d\bfsm{p}}{(2\pi)^3}%
 \tr_c\ln\left(1+L e^{-(E+\mu)/T}\right),
\ea \eeq where $E=\sqrt{p^2+M^2}$ with $M=m-\sigma$. As usual, the vacuum part of the above integral is divergent, so
we need to put the sharp cutoff $\Lambda$ in the momentum integral. We should note that in contrast to the $\mu=0$
case, $\Omega$ is not symmetric under the the gluonic charge conjugation $L\leftrightarrow L^+$~. \footnote{The
partition function is symmetric only under the full charge conjugation acting on both the quark and gluon sectors, \ie,
$L\to L^+$ together with $\mu\to-\mu$.}

Due to the global color $\mathrm{SU}(3)$ invariance of the thermodynamic
function of the loop, we can always find $g\in {\mathrm{SU}}(3)$ such
that $\Omega[L;\mu,T]=\Omega[g L_{\mathrm{diag}}g^{-1};\mu,T]%
\equiv\Omega[L_{\mathrm{diag}};\mu,T]$
where $L_{\mathrm{diag}}$ can be parameterized by
\beq
  L_{\rm
  diag}={\mathrm{diag}.}(e^{i\varphi_1},e^{i\varphi_2},e^{-i\varphi_1-i\varphi_2}).
\eeq
Then integrating out the ``phase'' $g$ using the property of the Haar
measure $dL=\mu(\varphi_1,\varphi_2) d\varphi_1 d\varphi_2[dg]$, the
partition function becomes
\beq
  Z[\mu,T]%
  =\int \mu(\varphi_1,\varphi_2)d\varphi_1 d\varphi_2%
   e^{-\beta V\Omega[L_{\mathrm{diag}};\mu,T]},
\eeq
where $\mu(\varphi_1,\varphi_2)$ is the measure for the eigenvalue
distribution of ${\mathrm{SU}}(3)$ that is proportional to the squared
Vandermonde determinant
$[\Delta(e^{i\varphi_1},e^{i\varphi_2},e^{-i\varphi_1-i\varphi_2})]^2$
\cite{Kogut:1981ez}.
The correctly normalized measure becomes
\beq
 \mu(\varphi_1,\varphi_2)=%
 \frac{(\sin(\varphi_1-\varphi_2)-\sin(2\varphi_1+\varphi_2)%
 +\sin(\varphi_1+2\varphi_2))^2}%
 {6\pi^2}~.
\eeq
Here the variables are defined in the interval
$\{\varphi_1,\varphi_2\}\in[0,2\pi]$. Now we introduce the independent
variables $\{\Phi,\bar\Phi\}$ by
$\Phi(\varphi_1,\varphi_2)=\frac{1}{3}\tr_c L_{\mathrm{diag}}$
and its conjugate $\bar\Phi=\Phi^*$; we have
$\mu(\varphi_1,\varphi_2)%
=\frac{9}{8\pi^2}(1-6\Phi\bar{\Phi}+4\Phi^3+4\bar\Phi^3%
-3\Phi^2\bar\Phi^2)$.
Then we may write down the full PNJL partition function as
\beq
\ba{rcl}
 Z_{\mathrm{PNJL}}[\mu,T]
   &\equiv&\int d\varphi_1 d\varphi_2
   e^{-S_\varphi[\Phi(\varphi_1,\varphi_2),%
   \bar\Phi(\varphi_1,\varphi_2);\mu,T]},\\[2ex]
   S_\varphi[\Phi,\bar\Phi,\sigma;\mu,T]&=&%
   \beta V(\Omega[\Phi,\bar\Phi,\sigma;\mu,T]%
   +V_{\mathrm{glue}}[\Phi,\bar\Phi])~.
\ea
\label{integral}
\eeq
In the above equation
$\Omega[\Phi,\bar\Phi,\sigma;\mu,T]=\Omega[L,\sigma;\mu,T]$
whose explicit form is given by
\beq
\ba{rcl}
 \Omega[\Phi,\bar\Phi,\sigma;\mu,T]&=&\frac{\sigma^2}{4G}%
  -2N_{\mathrm{f}}T\int\frac{d\bfsm p}{(2\pi)^3}%
 \ln(1+3\bar{\Phi}e^{-(E-\mu)/T}+3\Phi
   e^{-2(E-\mu)/T}+e^{-3(E-\mu)/T})\\
  &&-2N_{\mathrm{f}}T\int\frac{d\bfsm p}{(2\pi)^3}%
  \ln(1+3\Phi e^{-(E+\mu)/T}+3\bar{\Phi}
   e^{-2(E+\mu)/T}+e^{-3(E+\mu)/T})~.
\ea
\eeq
$V_{\mathrm{glue}}[\Phi,\bar\Phi]$
is the Polyakov-loop potential in the PNJL model specified below, which
includes the phenomenological mass term proportional to $\Phi^*\Phi$,
and the potential inspired by the group theoretical constraint, \ie, the
above-derived Vandermonde determinant.
\beq
 \frac{1}{T^4}V_{\mathrm{glue}}[\Phi,\bar\Phi]%
 =-\frac{b_2(T)}{2}\bar\Phi\Phi+b_4(T)%
 \ln(1-6\Phi\bar{\Phi}+4\Phi^3+4\bar\Phi^3%
   -3\Phi^2\bar\Phi^2).
\eeq $b_2(t)$ and $b_4(T)$ are the phenomenological parameters, and may
be set as in \cite{Hell:2008cc}.

In the next Section we show that the non-vanishing color density appears
in the model only at $\mu\ne 0$. The point is that $\Phi$ and $\bar\Phi$
are the {\it global} parameters in the PNJL model, and so are
$\{\varphi_1,\varphi_2\}$.
We should notice that by this simplification we can integrate over the
fermions in the partition function. This simplification, however,
inevitably introduces an artificial global color symmetry breaking in
this model, which may be regarded as an artifact as we discuss below.

\subsection{ A comment on the conjugated loop in the PNJL model.}
The expectation values for loop and conjugated loop can be determined in
principle by \cite{Dumitru:2005ng}
\beq
\ba{rcl}
 \langle\Phi\rangle_{\mu,T}&=&Z_{\mathrm{PNJL}}^{-1}%
  \int d\varphi_1 d\varphi_2%
 \left[(\Re\Phi)e^{-\Re S_\varphi}%
  \cos(\Im S_\varphi)%
 +(\Im\Phi)e^{-\Re S_\varphi}\sin(\Im
 S_\varphi)\right],\\
 \langle\bar\Phi\rangle_{\mu,T}&=&Z_{\mathrm{PNJL}}^{-1}%
  \int d\varphi_1 d\varphi_2%
 \left[(\Re\Phi)e^{-\Re S_\varphi}%
  \cos(\Im S_\varphi)%
 -(\Im\Phi)e^{-\Re S_\varphi}%
 \sin(\Im S_\varphi)\right].
\ea \eeq Here we have used the fact $\Im\Omega$ is odd with respect to the charge conjugation $\mathcal{C}$ acting on
gluon sector, \ie, $\Phi\leftrightarrow \bar\Phi$. It is notable that $\langle\Phi\rangle\ne\langle\bar\Phi\rangle$ but
both stay real \cite{Dumitru:2005ng}. Noting that the contribution to the $\sin(\Im S_\varphi)$ term in
$Z_{\mathrm{PNJL}}$ from $L^+$ configuration will be completely opposite to that from $L$, we could also write down the
partition function which is manifestly real as \beq
 Z_{\mathrm{PNJL}}[\mu,T]=\int d\varphi_1 d\varphi_2
 e^{-\Re S_\varphi}\cos(\Im S_\varphi).
\eeq It is often said that $\langle\Phi\rangle=\langle\bar\Phi\rangle$ in the mean field approximation, but this is not
true. It is the direct consequence of discarding the imaginary part of the action. In most of the literatures,
$\mathcal{C}$-odd piece $\Im S_\varphi$ is discarded, which is actually responsible for the equality between the
expectation values of loop and antiloop.

\section{Global color symmetry breaking in the PNJL model}
Now we turn to the problem of global color symmetry breaking. We adopt here the widely used simplification to discard
the imaginary piece of $S_\varphi$. In this way we should have $\langle\Phi\rangle=\langle\bar\Phi\rangle$ since the
simplified action no longer has the $\mathcal{C}$-odd part. This means that we should use the saddle point
approximation with a constraint $\Phi=\bar{\Phi}$. This can be done by taking a constraint
$\varphi_1=-\varphi_2\equiv\varphi$. This is similar to gauge fixing. Now we take the saddle point approximation,
namely we look for the values of $\Phi$ and $\sigma$ that minimize the thermodynamical potential: \beq
  \frac{\delta \Re S_\varphi}%
{\delta\varphi}\biggl|_{\{\varphi=\varphi_0,\sigma=\sigma_0\}}=0,\quad\frac{\delta \Re S_\varphi}%
{\delta\sigma}\biggl|_{\{\varphi=\varphi_0,\sigma=\sigma_0\}}=0,
\eeq
where $\Re S_\varphi=\beta V(\Omega[\Phi,\Phi,\sigma;\mu,T]%
+V_{\mathrm{glue}}[\Phi,\Phi])$ with $\Phi=\frac{1+2\cos\varphi}{3}$; the above equation defines the mean field values
of $\Phi_0$ and $\sigma_0$. This is the usual procedure in the PNJL model framework. Apparently this breaks the balance
between {\em red}, {\em green} quark and {\em blue} quark since the Wilson loop $L$ is asymmetric in color space within
the gauge in which $\varphi_1=-\varphi_2\equiv\varphi$. We name this configuration as $L_{\mathrm{diag}}^{(1)}$, that
is \beq
 L_{\mathrm{diag}}^{(1)}=\left(%
 \begin{array}{ccc}
 e^{i\varphi} & 0 & 0\\
 0 & e^{-i\varphi} & 0\\
 0 & 0 & 1\\
 \end{array}
 \right).
\eeq In this configuration we have observed a finite color density $\langle q^+ T_8q\rangle=n_8\ne0$ for $\mu\ne
0$~\cite{Abuki:2008ht}; the price to pay to avoid this fact is to introduce the color chemical potential $\mu_8$, whose
self consistently computed value assures the vanishing of the color charge (see also Ref.~\cite{GomezDumm:2008sk}).
\footnote{At $\mu=0$, quark density and antiquark density are balanced so that there is no color density in the system
even if $\mu_8=0$. This is consistent with the results obtained in \cite{Hell:2008cc}.} When the diquark condensate
comes to play a role, $\langle q^+T_8q\rangle=n_8\ne 0$ at $\mu_8=0$ regardless of the presence/absence of the Polyakov
loop. One could think that it does not bring any problem because of a residual discrete ``gauge'' freedom which remains
even after a continuous gauge freedom is lost by the gauge fixing $L\to gL_{\mathrm{diag}}g^{-1}$. This remnant is
nothing but a permutation of $(r,g,b)$ quarks.

One must, of course, keep in mind that several configurations that give the same $\Phi_0$ may contribute to the total
$Z_{\mathrm{PNJL}}^{\mathrm{MF}}[\Phi_0,\sigma_0;\mu,T]$ via the integral given in Eq.~(\ref{integral}). Let us first
introduce explicitly the elements of the permutation group, which is subgroup of $\mathrm{SU}(3)$. The followings are
the elements of (1)$\Leftrightarrow(r,g ~flip )$, (2)$\Leftrightarrow(g,b~ flip)$, (3)$\Leftrightarrow (b,r ~flip)$.
\beq \ba{rcl}
g_{rg}=\left(%
\begin{array}{ccc}
0&1&0\\
1&0&0\\
0&0&1\\
\end{array}%
\right),\quad g_{gb}=\left(%
\begin{array}{ccc}
1&0&0\\
0&0&1\\
0&1&0\\
\end{array}%
\right),\quad g_{br}=\left(%
\begin{array}{ccc}
0&0&1\\
0&1&0\\
1&0&0\\
\end{array}%
\right)
\ea
\eeq
Using the above discrete transformations, we can produce another
configuration (different gauge) which realizes the same value of
$\Phi$.
We name the $(r,g)$ flipped configuration $L^{(2)}_{\mathrm{diag}}$,
that is
\beq
 L_{\mathrm{diag}}^{(2)}=%
 g_{rg}L_{\mathrm{diag}}g_{rg}^{-1}=\left(%
 \begin{array}{ccc}
  e^{-i\varphi} & 0 & 0\\
  0 & e^{i\varphi} & 0\\
  0 & 0 & 1\\
 \end{array}
 \right).
\eeq From these $L_{\mathrm{diag}}^{(1)}$ and $L_{\mathrm{diag}}^{(2)}$, we can construct four more configurations as
\beq \ba{rcl}
 L_{\mathrm{diag}}^{(3)}=g_{gb}L_{\mathrm{diag}}^{(1)}g_{gb}^{-1}%
 &=&\left(%
 \begin{array}{ccc}
  e^{i\varphi} & 0 & 0\\
  0 & 1 & 0\\
  0 & 0 & e^{-i\varphi}\\
 \end{array}\right), %
 \quad L_{\mathrm{diag}}^{(4)}=g_{gb}L_{\mathrm{diag}}^{(2)}g_{gb}^{-1}%
 =\left(%
 \begin{array}{ccc}
  e^{-i\varphi} & 0 & 0\\
  0 & 1 & 0\\
  0 & 0 & e^{i\varphi}\\
 \end{array}\right),\\
 L_{\mathrm{diag}}^{(5)}=g_{br}L_{\mathrm{diag}}^{(1)}g_{br}^{-1}%
 &=&\left(%
 \begin{array}{ccc}
  1 & 0 & 0\\
  0 & e^{-i\varphi} & 0\\
  0 & 0 & e^{i\varphi}\\
 \end{array}\right), %
 \quad L_{\mathrm{diag}}^{(6)}=g_{br}L_{\mathrm{diag}}^{(2)}g_{br}^{-1}%
 =\left(%
 \begin{array}{ccc}
  1 & 0 & 0\\
  0 & e^{i\varphi} & 0\\
  0 & 0 & e^{-i\varphi}\\
 \end{array}\right).
\ea
\eeq
All these configurations $\{L^{(1)},L^{(2)},\cdots,L^{(6)}\}$ give the
same $\Phi$. In the original variables $(\varphi_1,\varphi_2)$
parametrizing $L_{\mathrm{diag}}$ as $L_{\mathrm{diag}}=\mathrm{diag.}%
(e^{i\varphi_1},e^{i\varphi_2},e^{-i\varphi_1-i\varphi_2})$,
these configurations correspond to
$\{(\varphi,2\pi-\varphi)_1,(2\pi-\varphi,\varphi)_2,(\varphi,0)_3,%
(2\pi-\varphi,0)_4,(0,\varphi)_5,(0,2\pi-\varphi)_6\}$. This will become clearer in the diagonal color $\mathrm{SU}(3)$
basis, $\{\lambda_3,\lambda_8\}$, writing $L_{\mathrm{diag}}=e^{i\phi_3\lambda_3+i\phi_8\lambda_8}$. In this base,
these configurations make a hexahedron in the color $(3,8)$ plane, \ie,
$(\phi_3,\phi_8)=\{(\varphi,0)_1,(-\varphi,0)_2,%
(\frac{\varphi}{2},\frac{\sqrt{3}}{2}\varphi)_3,%
(-\frac{\varphi}{2},-\frac{\sqrt{3}}{2}\varphi)_4,%
(\frac{\varphi}{2},-\frac{\sqrt{3}}{2}\varphi)_5,%
(-\frac{\varphi}{2},\frac{\sqrt{3}}{2}\varphi)_6\}$. It is worth to note that each configuration has a different color
charge although all these configurations are degenerate in energy. Let us denote the color densities in configuration
(1) generally as $(n_3,n_8)_1$ even though $n_3=0$ in this case. Then, noting that color density transforms by each
discrete gauge transformation as $\langle q^+T_{\alpha}q\rangle\to\langle q^+U^+_{ab}T_\alpha U_{ab}q\rangle$, we
obtain color density in each configuration
$(n_{3(i)},n_{8(i)})=\{(n_3,n_8)_1,(-n_3,n_8)_2,%
(\frac{n_3}{2}+\frac{\sqrt{3}}{2}n_8,\frac{\sqrt{3}}{2}n_3-\frac{n_8}{2})_3,%
(-\frac{n_3}{2}+\frac{\sqrt{3}}{2}n_8,-\frac{\sqrt{3}}{2}n_3-\frac{n_8}{2})_4,%
(\frac{n_3}{2}-\frac{\sqrt{3}}{2}n_8,-\frac{\sqrt{3}}{2}n_3-\frac{n_8}{2})_5,%
(-\frac{n_3}{2}-\frac{\sqrt{3}}{2}n_8,\frac{\sqrt{3}}{2}n_3-\frac{n_8}{2})_6\}$.
We notice that $n_{3(i)}^2+n_{8(i)}^2=n_3^2+n_8^2$ is a gauge invariant
quantity as it should \cite{Buballa:2005bv}.

One could argue that the average of color densities over all these configuration,
$\frac{1}{6}\sum_{i=1}^6(n_3,n_8)_{i}=(0,0)$ so that we would not need to worry about the color density. Of course this
is not the case. The reason is that the thermodynamic potential of the system is not the average of those, that is
$\frac{1}{6}\sum_{i=1}^6\Omega_{(i)}$ where $\Omega_{(i)}$ is the thermodynamic potential for the configuration
$(\phi_3,\phi_8)_i$, \ie, $\beta V\Omega_{(i)}=\Re S_\varphi|_{(\phi_3,\phi_8)_i}$. To see this more clearly, we write
the mean-field approximated partition function as \beq \ba{rcl}
  Z_{\mathrm{PNJL}}^{\mathrm{MF}}[\Phi_0,\sigma_0;\mu,T]%
  \equiv e^{-\beta V\Omega_{\mathrm{PNJL}}^{\mathrm{MF}}%
  [\Phi_0,\sigma_0;\mu,T]}=\sum_{(\phi_3,\phi_8)_i}%
  e^{-\beta V\Omega_{(i)}},
\ea
\eeq
thus we have
\beq
  \Omega_{\mathrm{PNJL}}^{\mathrm{MF}}[\Phi_0,\sigma_0;\mu,T]=-\frac{T}{V}\ln\left(\sum_i e^{-\beta
  V\Omega_{(i)}}\right)\stackrel{V\to\infty}{\to}%
  \min\big(\Omega_{(1)},\Omega_{(2)},\cdots,\Omega_{(6)}\big).
\eeq
Of course, in current case with no diquark condensate, each potential is
degenerate in energy although each has a different color
charge. Nevertheless  there is no valid argument to choose the average
of $\{\Omega_{(i)}\}$ as a definition of
$\Omega^{\mathrm{MF}}_{\mathrm{PNJL}}$.
Furthermore, since the color symmetry is realized as a global symmetry
in the PNJL model, these degenerated vacua are extremely sensitive to
the infinitesimal external field which tends to align the vacuum to a
specific direction, that is a famous spontaneous symmetry
breakdown \cite{Nambu:1961tp}.
This can be easily illustrated as follows. If we introduce an
infinitesimal external field $\epsilon_1$ which tries to align the
vacuum to $\Omega^{(1)}$ out of six directions, then we have the vacuum
alignment to the $\Omega^{(1)}$ direction when the thermodynamic limit
is taken as $V\to\infty$. Then after we switch off the external field
$\epsilon_1\to 0$, the vacuum remains in the $(1)$ direction. That is
\beq
\Omega_{\mathrm{PNJL}}^{\mathrm{MF}}[\Phi_0,\sigma_0;\mu,T]%
  =-\lim_{\epsilon_1\to0}\lim_{V\to\infty}%
  \frac{T}{V}\ln\left(\sum_i e^{-\beta
  V\Omega_{(i)}}\right)=\Omega_{(1)}.
\eeq In this way, the global color symmetry can be broken in the adopted mean field PNJL model, and as a consequence,
nonvanishing color charge remains in the system. Whichever gauge is chosen among six configurations, the statement that
color charge does not vanish is gauge invariant since $\sqrt{n_3^2+n_8^2}$ is gauge invariant. Of course, we think that
this global color symmetry breaking in the PNJL model is an artifact of the model, possibly of the adopted mean field
approximation.

\begin{figure}[t]
\centerline{
\includegraphics[width=8cm]{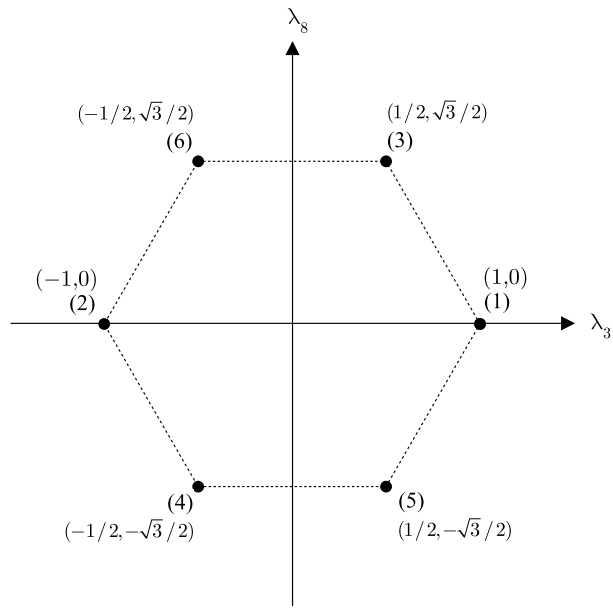}
}
\caption{The discrete gauge configurations of the Wilson line
 $L(\phi_3,\phi_8)=e^{i\phi_3\lambda_3+i\phi_8\lambda_8}$
 in the $(\lambda_3,\lambda_8)$ plane which result in the same value of
 $\Phi=\frac{1}{3}\tr_c L$.
 The axis scale is normalized by $\varphi$ such that
 $\Phi=(1+2\cos\varphi)/3$.
 Each configuration yields a different color charge density;
 Denoting the color charge density of configuration (1), as $(n_3,n_8)$,
 then we have $(-n_3,n_8)$ for configuration (2),
 $(\frac{n_3}{2}+\frac{\sqrt{3}n_8}{2},\frac{\sqrt{3}n_3}{2}-\frac{n_8}{2})$
 for (3),
 $(-\frac{n_3}{2}+\frac{\sqrt{3}n_8}{2},-\frac{\sqrt{3}n_3}{2}-\frac{n_8}{2})$
 for (4),
 $(\frac{n_3}{2}-\frac{\sqrt{3}n_8}{2},-\frac{\sqrt{3}n_3}{2}-\frac{n_8}{2})$
 for (5),
 $(-\frac{n_3}{2}-\frac{\sqrt{3}n_8}{2},\frac{\sqrt{3}n_3}{2}-\frac{n_8}{2})$
 for (6).
 The value of $n_3^2+n_8^2$ is the same for all configurations, and thus
 is gauge invariant.
}
\label{fig:1}
\end{figure}

One may still think that this kind of problem does not appear if we take as a dynamical variable $\Phi$ itself rather
than the gauge field $\{\phi_3,\phi_8\}$. In this case one can not compute each color density, so we cannot judge
whether the system is color neutral or not. Furthermore this does not work as soon as we take the diquark condensation
into account. In that case, we need to treat $\{\phi_3,\phi_8\}$ as the dynamical variable since the quark part of
potential is no longer function of a gauge invariant quantities, $\Phi$ and $\bar\Phi$. Moreover, colored diquarks work
as external fields which prefer some specific direction of vacuum among six configurations. When the critical
temperature is approached, this external field vanishes rapidly, $\Delta\to 0$ as $T-T_c\to 0^-$, but the vacuum
alignment remains in the system. In this way, we can not avoid the problem of color symmetry breaking at finite
density, $\mu\ne0$.

Before going on let us make a clarification regarding the breaking of color symmetry in the PNJL model. It is
well-known that the a local gauge symmetry can never be broken as stated in the Elitzur's theorem
\cite{Elitzur:1975im}. Thus the expectation value of an operator which is not invariant under the gauge transformation
should vanish after all the configurations of gauge fields are properly taken into account into the partition function.
Nevertheless, it is sometimes useful to fix a gauge and define a gauge dependent order parameter in weak coupling
perturbative calculations just to extract gauge independent physical quantities, as done in the BCS theory of
superconductivity. In the BCS theory, one assumes a non-vanishing electron pair condensate which leads to a physical
gap in a electron spectrum.

This is what happens in the PNJL model, as well as in the NJL model applied to color superconductive phases (see next
Section). In the PNJL model within the adopted mean field approximation, one usually introduces a gauge dependent
matrix-valued order parameter $L$ (working in the Polyakov gauge), and then he tries to look at the gauge-independent
Polyakov-loop $\Phi=\frac{1}{3}\tr_c L$ which is, in a pure gauge limit, related to the free energy of a single static
quark excitation in a gluonic heatbath. A color-symmetry breaking in a PNJL model might be understood in this way. But
this scenario immediately brings the trouble we have encountered in this section, that the gauge-independent squared
sum of color densities is nonvanishing at $\mu\ne0$  unless either $\Phi\ne 1$ or $T=0$. After all, we conclude that
this color-symmetry breaking in a PNJL model is an artifact of too naive saddle-point approximation for the gauge
dependent quantity $L$. In order to overcome this problem, we need to somehow construct the gauge-independent
approximation scheme.

\section{Color superconductivity within the PNJL model}
Let us finally discuss some difficulty to tackle the problem of color superconductivity in the PNJL model, although
this problem is studied to some extent but with an {\em implicit} assumption which makes the problem much simpler
\cite{Roessner:2006xn,Abuki:2008ht,Ciminale:2007sr}. We wish here to clarify what is the problem behind these studies,
as we already mentioned in \cite{Abuki:2008ht}. Instead of Eq. (\ref{lag}), we start with a matrix model combined with
the NJL model described by the following Euclidean lagrangian, which would lead the system to be in the 2SC
superconducting state \cite{Rapp:1997zu,Alford:1997zt} at high density. \beq
 \mathcal{L}_E(q,\bar{q},A_4)=q^+(\gamma_0(\partial_\tau+i A_4)%
  +i\bfm{\gamma}%
  \cdot{\nabla}-\ds{\mu})q-\frac{G_D}{4}\left|%
  q^tC\gamma_5\bfm{\ep}_{a}^{\mathrm{(c)}}%
  \bfm{\ep}^{\mathrm{(f)}}q\right|^2.
\eeq We introduced $\bfm{\ep}_a^{\mathrm{(c)}}$ and $\bfm{\ep}^{\mathrm{(c)}}$, the antisymmetric matrix in color and
flavor space respectively, \ie, $[\bfm{\ep}_a^{\mathrm{(c)}}]_{bc}=\ep_{abc}$ and
$[\bfm{\ep}^{\mathrm{(f)}}]_{ij}=\epsilon_{ij}$. Introducing the bosonic field
$\Delta_a\sim\frac{G_d}{2}q^tC\gamma_5\bfm{\ep}^{(c)}_a\bfm{\ep}^{(f)}q$, and integrating out fermions results in the
expression of the partition function in terms of a {\em global} matrix model \beq \ba{rcl}
 Z[\mu,T]&=&\int_{\mathrm{P.B.}}\Pi_x\Pi_{a=1}^3d\Delta_a(x)
 d\bar{\Delta}_a(x)e^{-S_{\mathrm{eff}}[\Delta_a(x),\bar\Delta_a(x)]}\\
 e^{-S_{\mathrm{eff}}[\Delta_a,\bar\Delta_a]}%
 &=&\int dL e^{-\sum_a\int d\tau d\bfsm{x}%
  \frac{\Delta_a^*(x)\Delta_a(x)}{G_D}%
  +\frac{1}{2}\log\mathrm{Det}\left[\left(\begin{array}{cc}
 i\gamma^0\ds{D}[\mu,A_4] & -\Delta_a(x)\gamma_5\bfm{\ep}_{a}^{\mathrm{(c)}}%
   \bfm{\ep}^{\mathrm{(f)}}\\
-\Delta_a(x)\gamma_5\bfm{\ep}_{a}^{\mathrm{(c)}}\bfm{\ep}^{\mathrm{(f)}}&
    i\ds{D}[-\mu,-A_4]^t\gamma^0 \\
 \end{array}
 \right)\delta_{\mathrm{P}}(x-y)\right]},
\ea
\eeq
where the covariant derivative in the Euclidean space is defined by
$i\ds{D}[\mu,A_4]=-\gamma^0\partial_\tau+i\gamma^i\partial_i%
+(\mu-iA_4)\gamma^0$.
Here $A_4=\sum_{\alpha=1}^{8}\phi_\alpha\lambda_\alpha$ is the temporal
background gauge field related to the Wilson line by $L=e^{i\beta A_4}$,
$\delta_{\mathrm{P}}(x-y)$
is again the delta function periodic in $\tau$.  The transpose operation
${}^t$
should be supposed to work only in color and flavor space, not in the
Dirac space.

We adopt the saddle point approximation for $S_{\mathrm{eff}}$ and make the following homogeneous ansatz for the
superconducting condensate, $\Delta_a(x)=\Delta\delta_{a3}$ with $\Delta$ being real, obtaining a matrix model inspired
PNJL action \beq
 e^{-S_{\mathrm{PNJL}}[\vec{\Delta}^*\cdot\Delta]}%
 =\int dLe^{-S_\varphi[\vec{\Delta},A_4;\mu,T]},
\eeq where $S_\varphi=\beta V\Omega[\vec{\Delta},A_4;\mu,T]$ whose expression will be given below, and we have
introduced the condensate vector $\vec{\Delta}=(0,0,\Delta)$ in the color antitriplet space. We have taken care the
fact that when integrated over $L$, the action $S_{\mathrm{eff}}$ is only a function of the gauge invariant convolution
$\vec{\Delta}^*\cdot\Delta$, while it is not before the integration, \ie, $S_\varphi$, or equivalently $\Omega$ depend
on the relative angle of $L$ and $\vec{\Delta}$ in the color space as we shall see later. The real part of the
thermodynamic potential in the presence of a constant background gauge $A_4$ can be calculated by \footnote{The readers
are refereed to \cite{Abuki:2008ht,Abuki:2005ms} for more detail of the derivation.} \beq
 \Re\Omega[\vec{\Delta},A_4;\mu,T]%
 =\frac{\Delta^2}{G_D}-\sum_{A=1}^{24}\int%
 \frac{d\bfm{p}}{(2\pi)^3}\left[\Re E_A(p)+2T\ln|\!|%
 1+e^{-E_A(p)/T}|\!|\right].
\eeq
${E_A}(p)$
is chosen from the eigenvalues of the following non-hermitian 48 by 48
hamiltonian density matrix such that $\Re E_{A}$ is
positive.\footnote{This is always possible since the eigenvalues of the
Hamiltonian density appears in sets of eigenvalues, $\{E_A,-E_A\}$
because of the Nambu-Gorkov degeneracy.}
The index $A$ refers to the quasiquark degrees of freedom that can be
enumerated as
$3\mathrm{(color)}\times2\mathrm{(flavor)}\times4\mathrm{(Dirac)}$.
\beq
 {\mathcal H}(p)=\left(%
 \begin{array}{cc}
   \alpha\cdot\bfm{p}-(\mu-iA_4)&
    \Delta\gamma_5\bfm{\ep}^{\mathrm{(c)}}_3\bfm{\ep}^{\mathrm{(f)}}\\
   \Delta\gamma_5\bfm{\ep}^{\mathrm{(c)}}_3\bfm{\ep}^{\mathrm{(f)}}%
   & \alpha\cdot\bfm{p}+(\mu-iA_4)\\
 \end{array}
 \right),
\eeq where $\alpha^i=\gamma^0\gamma^i$. It is important to note now that the direction of color symmetry breaking is
gauge fixed to the third direction in color space.

With this fact in mind, let us see what happens when we wish to
integrate in the $\mathrm{SU}(3)$ matrix $L$. We again make use of the
polar/phase decomposition of $L$ as
\beq
L=g^{-1}L_{\rm diag}g,\quad dL=\mu(\varphi_1,\varphi_2)d\varphi_1 %
   d\varphi_2[dg]~.
\eeq
Ignoring the imaginary part of the PNJL model action as usual, the
partition function becomes
\beq
\ba{rcl}
 Z_{\mathrm{PNJL}}(\mu,T)&=&\int
 \mu(\varphi_1,\varphi_2)d\varphi_1d\varphi_2%
 [dg]e^{-\Re S_\varphi%
 [\vec{\Delta},\,g^{-1}A_4^{\mathrm{diag}}g;\,\mu,T]}\\
 &=&\int
 \mu(\varphi_1,\varphi_2)d\varphi_1d\varphi_2%
 [dg]e^{-\Re S_\varphi%
 [g^*\vec{\Delta},\,A_4^{\mathrm{diag}};\,\mu,T]},
\ea
\eeq
where $g^*$ is the complex conjugate of the phase matrix $g$.
Here we clearly see that the phase $g$ does not disappear from the
integrand so that we can not trivially perform the integration over the
phase. This means that the saddle-point solutions to $Z_{\mathrm{PNJL}}$
with respect to the gauge fields $\{\varphi_1,\varphi_2\}$ clearly depend
on the relative angle $g^*$ between $A_4$ and $\vec{\Delta}$.
This is in contrast to the case where we do not have the diquark
condensate in the system. This means that we can no longer restrict
ourselves to diagonal form of the temporal gauge $A_4$ when the
orientation of diquark condensate is fixed to the third axis of the
color space. Otherwise, we can not limit ourselves to the assumption
$\vec{\Delta}=(0,0,\Delta)$
when $A_4$ is parametrized by a diagonal form. This fact is already
stressed in \cite{Abuki:2008ht}, but it is usually assumed in the
literatures that both $A_4$ is diagonal and
$\Delta=(0,0,\Delta)$~\cite{Roessner:2006xn}.

What happens in that case? Let us assume without justification that the
integration over $[dg]$ can be done
\beq
\ba{rcl}
 Z_{\mathrm{PNJL}}(\mu,T)&=&\int
 \mu(\varphi_1,\varphi_2)d\varphi_1d\varphi_2%
 e^{-\beta V\Re\Omega%
 [\vec{\Delta},A_4^{\mathrm{diag}};\mu,T]},
\ea
\eeq
where $\vec{\Delta}=(0,0,\Delta)$. Then we work the saddle point
approximation in $\varphi_1$ and $\varphi_2$. Again restricting $\Phi$
to be real, letting $\Phi=\frac{1+2\cos\varphi}{3}$, the partition
function becomes
\beq
 Z^{\mathrm{MF}}_{\mathrm{PNJL}}(\mu,T)=\sum_{\{\phi_3,\phi_8\}_{i}}%
 e^{-\beta V\big[\Re\Omega%
 [\vec{\Delta},A_{4(i)}^{\mathrm{diag}};\mu,T]%
 +V_{\mathrm{glue}}[\Phi,\Phi]\big]}.
\eeq $A_{4(i)}^{\mathrm{diag}}$ is the temporal background gauge field in the configuration $\{\phi_3,\phi_8\}_i$. We
have exponentiated the Vandermonde determinant and introduced a phenomenological Polyakov loop potential as before. In
this case, the degeneracy in six discrete gauge configurations is lifted away. We have two degenerated minima in this
case, that is configuration (1) and (2). This is a kind of vacuum alignment, because only the configuration (1) and (2)
do not bring any additional symmetry breaking in $\Re\Omega$ but the original one, \ie,
$\mathrm{SU}(3)_{\mathrm{c}}\to\mathrm{SU}(2)_{\mathrm{c}}$ caused by diquark condensate $\vec{\Delta}=(0,0,\Delta)$.
In the thermodynamic limit $V\to\infty$, the state should be selected to be either in configuration (1) or (2) again
not to their average. Even though both states in this case turn out to have the same color charge density $(0,n_8)$
because of vanishing $n_3$, the physical state cannot be the statistical (symmetric) average of these two states
because the PNJL model has only a global color symmetry, and the quantum field theory has an infinite number of degrees
of freedom in the limit $V\to\infty$. This is in contrast to quantum mechanics where we can construct the symmetric
states by average of two states. When temperature is increased, the 2SC state eventually undergoes a 2nd order
transition to the normal quark matter phase. $\Delta$ decreases to zero when the critical temperature is approached,
but the alignment of $A_4$ in the color space remains in (1) or (2) configuration, both of which yield nonvanishing
color density $n_8\ne0$ even in the normal quark matter phase at $\mu\ne0$. We thus conclude that the PNJL model
without color chemical potentials at finite $\mu\ne0$ is not free from appearance of color charge density even if
diquark fields are absent.

\section{Conclusions}
We have discussed how color neutrality should be implemented in the adopted mean field PNJL model at finite density. We
have given detailed formal arguments that explain why we need to introduce a color chemical potential at $\mu\neq0$ in
order to assure color neutrality of the system. We have compared our results with those obtained in
Ref.~\cite{Hell:2008cc}, showing that there is not discrepancy at $\mu=0$ (which is the regime investigated numerically
in the aforementioned paper).

Moreover we have pointed out that the mean field solution can not be
computed as an average over the degenerate ground states: as a matter of
fact, spontaneous symmetry breaking means that one can choose only one of
the degenerate ground states.
% moreover in the case of the broken descrete
%symmetry, which is relevant in the present context, there are no
%Goldstone bosons that interpolate among the various degenerate solutions
%since the broken symmetry is a discrete one.
In fact, the various ground states of the PNJL model can not communicate
among them because they construct different Hilbert/Fock spaces, \ie, the
overlap between those spaces vanishes after $V\to\infty$; we should
choose only one of them for making calculations. This procedure, which is
common both to discrete and to continuous symmetry, is very popular in
statistical mechanics, when one for example studies effective models of
spontaneous magnetization of Weiss domains: Nature chooses only one
direction for magnetization, even if the ground state possesses an
infinity of degenerate ground states corresponding to the infinite
possible orientation of the magnetization.

We have also discussed a subtle point related to the simultaneous diagonalization of the Polyakov loop $L$ and the
color superconductive gap matrix $\Delta$. We have shown that in principle it is not correct to take both of them in a
diagonal form, since one needs two color rotations in order to achieve this: the color rotation that makes $L$ diagonal
does not necessarily make $\Delta$ diagonal and vice versa. Thus in principle one should work either with a diagonal
$L$ or with a diagonal $\Delta$. Furthermore we have noticed that, defining a non superconductive state as a limit for
$\Delta\rightarrow 0$ of a superconductive one, the vacuum alignment makes both mathematically and physically
impossible to define the ground state as a mean of the several degenerate ground states of the model. As a consequence,
we can not avoid non-vanishing color charge density even in the non superconducting phase without introducing
appropriate color chemical potentials, at least in the mean field approximation adopted here.

\end{document}